Submitted to Physica B (2012)

Positron Studies of Defects (PSD-11)

# Study of ageing in Al-Mg-Si alloys by positron annihilation spectroscopy

J. Banhart[a,b] *, M. Liu[b], Y. Yong[b], Z. Liang[b], C.S.T. Chang[b], M. Elsayed[c], M.D.H. Lay[d]

[a]*Berlin Institute of Technology (TU Berlin), Institute of Materials Science and Technology, Hardenbergstr. 36, 10623 Berlin, Germany*
[b]*Institute of Applied Materials, Helmholtz Centre Berlin, Hahn-Meitner Platz 1, 14109 Berlin, Germany*
[c]*Martin-Luther-University Halle-Wittenberg, Von-Danckelmann-Platz 3, 06120 Halle, Germany*
[c]*CSIRO Materials Science and Engineering, Clayton, Victoria 3169, Australia*

**Abstract**

In many common Al-Mg-Si alloys (6000 series) intermediate storage at or near ‘room temperature’ after solutionising leads to pronounced changes of the precipitation kinetics during the ensuing artificial ageing step at ≈180°C. This is not only an annoyance in production, but also a challenge for researchers. We studied the kinetics of natural ‘room temperature’ ageing (NA) in Al-Mg-Si alloys by means of various different techniques, namely electrical resistivity and hardness measurement, thermoanalysis and positron lifetime and Doppler broadening (DB) spectroscopy to identify the stages in which the negative effect of NA on artificial ageing might appear. Positron lifetime measurements were carried out in a fast mode, allowing us to measure average lifetimes in below 1 minute. DB measurements were carried out with a single detector and a $^{68}$Ge positron source by employing high momentum analysis. The various measurements show that NA is much more complex than anticipated and at least four different stages can be distinguished. The nature of these stages cannot be given with certainty, but a possible sequence includes vacancy diffusion to individual solute atoms, nucleation of solute clusters, Mg agglomeration to clusters and coarsening or ordering of such clusters. Positron lifetime measurements after more complex ageing treatments involving storage at 0°C, 20°C and 180°C have also been carried out and help to understand the mechanisms involved.

## 1. Introduction

Pure aluminium is light but soft. Hardening (strengthening) of aluminium is therefore a prerequisite for most structural applications. Alloying can lead to large strength increases especially when the technique of *age hardening* is applied that comprises preparation of a non-equilibrium state by first quenching the alloy from the single-phase region to low temperature and then allowing the system to move towards equilibrium by applying a sequence of ageing steps that lead to the formation of precipitates. Alloys based on the ternary system Al-Mg-Si with additions of Cu, Fe, Mn, Cr, etc. are very common. The precipitation sequence in such alloys is usually described as: SSSS→clusters→β''→β'→β, where the supersaturated solid solution is rich in non-equilibrium vacancies that enable a passage through a series of metastable precipitates towards the equilibrium phase. A sequence of precipitation has been discussed in many variants [1] but none of the proposed sequences can explain all the phenomena observed experimentally. There are various reasons for this: first, the exact sequence depends on temperature, alloy composition and possibly quenching conditions. Especially when industrial alloys are studied, the presence of Fe, Mn, Cu etc. complicates the situation. Second, a linear sequence appears to be an oversimplification In reality, different precipitation processes compete with each other already in the clustering stage [2]. Third, industrial ageing involves more than one temperature. After quenching, alloys are often kept at 'room temperature' for a while and are processed there before they are aged at temperatures around 180°C. This 'natural' pre-ageing has a pronounced effect on the subsequent 'artificial ageing' step, often a negative one [3]. A simple linear precipitation chain cannot explain such behaviour. It is therefore essential to study these processes both isothermally at different temperatures and also, in a second step, during step ageing at different temperatures.

* Corresponding author. Tel.: +49-30-8062-42710; fax: +49-30-8062-42079.
*E-mail address*:john.banhart@tu-berlin.de.

Isothermal studies of NA (at ~20°C) reveal that NA is not a continuous process. Hardness, electrical resistivity and heat evolution show distinct stages. This is illustrated for a pure ternary model alloy in Fig. 1. Hardness evolves in at least two stages, electrical resistivity in at least three, whereas the thermal signal of clustering can be subdivided into at least two stages. Some of the transition times between various stages as measured by different methods show a rough coincidence. In stage I, resistivity varies linearly and the negative effect on AA develops (not shown here) within about 10 minutes. In stage II, both resistivity and hardness increase rapidly for about an hour, during which a first stage of cluster formation (C0,C1) is detected by DSC. In stage III, the rates of both hardening and electrical resistivity increase level off and a second clustering mechanism (C2) starts to dominate DSC. These stages vary with alloy composition and are dominated by the amount of solute (Mg,Si) present [4]. There might even be a stage IV but this cannot be identified with certainty from the available data.

Positron annihilation spectroscopy (PAS) is sensitive to vacancies that play an important role in age hardening. Therefore, monitoring NA with PAS is near at hand. Such studies have been carried out for Al-Cu-Mg and Al-Mg-Zn alloys, see Ref. [5], but with data acquisition times in the order of 30 min, which in view of Fig. 1 is insufficient to resolve the changes taking place. We evaluated whether the power to resolve temporal changes can be improved and apply PAS to NA of Al-Mg-Si alloys. One aim is to reproduce the stages identified in Fig. 1 by the PAS signal.

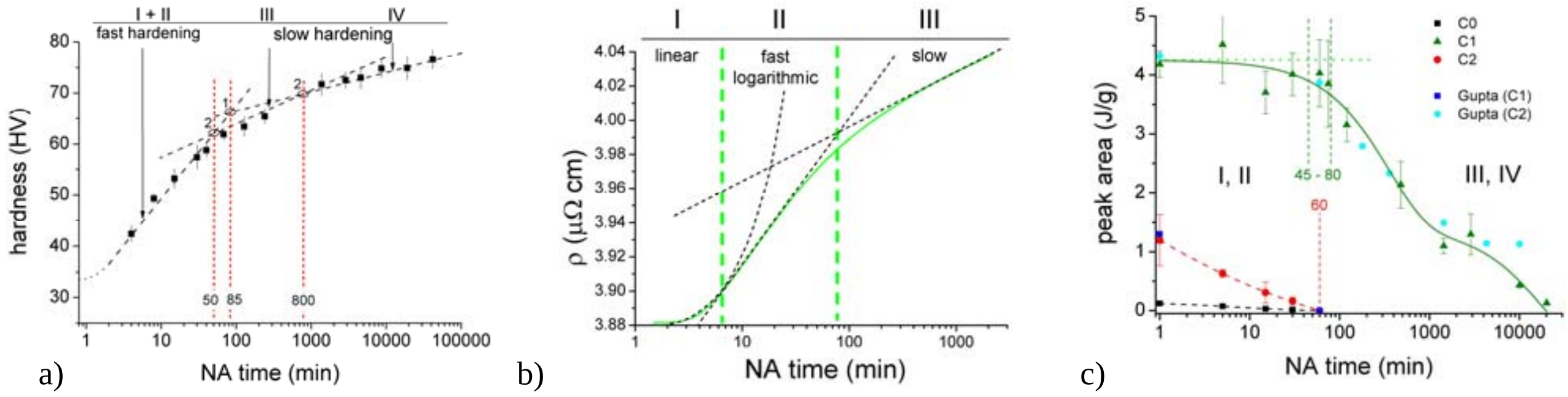

Fig. 1. Time evolution of a) hardness [3], b) electrical resistivity, c) heat release [2], all at 'room temperature'. Alloy was Al-0.6wt.%Mg-0.8wt.% Si (6-8) after solutionising at 535°C for 30 min and quenching into ice water. Roman numerals denote stages of NA (following [6]).

## 2. Materials and Methods

### *2.1. Materials*

All alloys studied were prepared by Hydro Aluminium Bonn from pure elements as described in Ref. [6]. In this study, we present measurements on pure ternary alloys which we shall specify by their nominal composition in wt.% in the format **x–y** = Al - x/10 wt.%Mg - y/10 wt.% Si. For example, '4-4' means Al-0.4wt.%Mg-0.4wt.% Si.

### *2.2. Positron Annihilation Lifetime Spectroscopy (PALS)*

Most measurements were carried out with an Ortec PALS system at CSIRO in Melbourne, see description in Ref. [6]. In order to check the results, experiments were also carried out with a PALS system at the University of Halle and a system built at the University of Bonn, now operated at the Helmholtz Centre Berlin.

The first objective of this study was to optimise the acquisition time of PALS spectra. Our ansatz is based on the observation that positron spectra in 6000 alloys after quenching and during NA are well described by one lifetime component, in particular no component related to bulk aluminium (≤0.160 ns) appears clearly. Therefore, it is not necessary to collect a large amount of data since a reliable estimate for an average lifetime can be obtained from small data sets as illustrated in Fig. 2: The lifetimes derived from many individual spectra measured with low statistics scatter around the lifetime obtained using a single spectrum containing $7\times10^7$ counts. The scatter of all lifetimes is comparable to the statistical error of that single lifetime fit. Therefore, for the purpose of detecting positron lifetime *changes*, low statistics are sufficient.

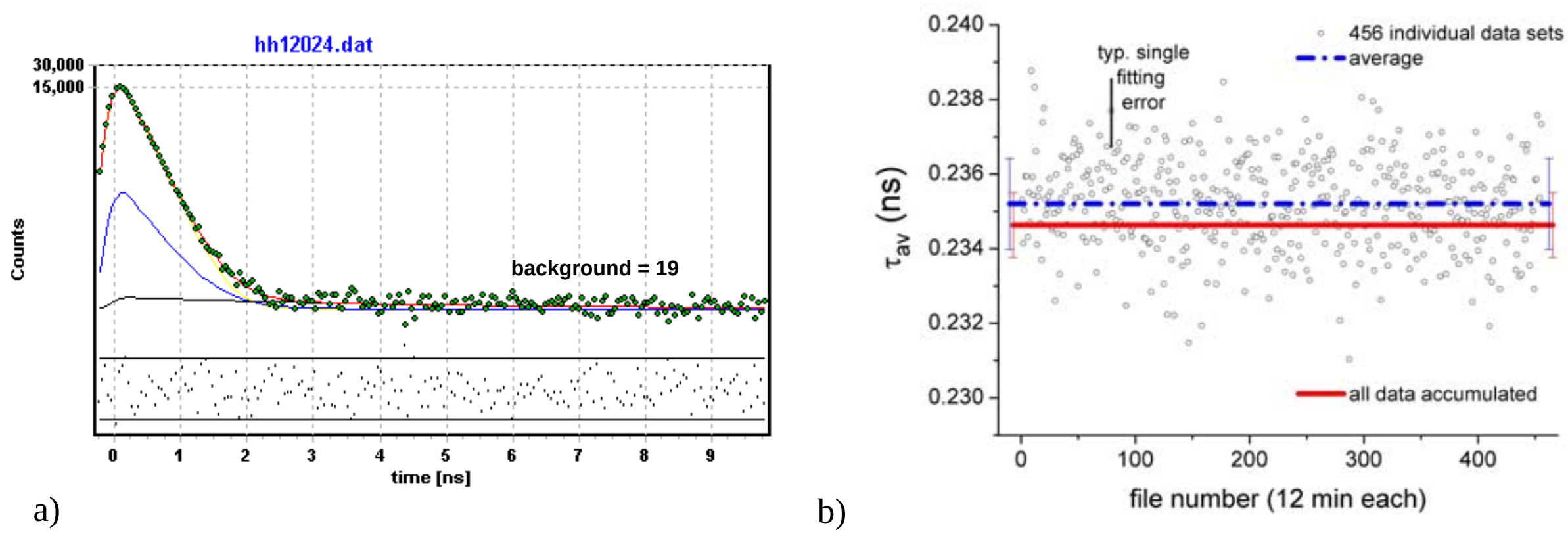

Fig. 2. Positron lifetime measurements on alloy 4-4 after solutionising and quenching. Measurement carried out at –50° in a cryo sample holder after ≈2 min of processing at ‘room temperature’. 456 spectra were recorded, each containing 155,000 counts (12 min acquisition time, 15,000 peak counts). a) fit of single exponential decay to one such spectrum (red line). Source contributions (blue and black line) and fitting residuals are also shown. b) lifetimes determined for all spectra (circles) and their average are shown (dashed line). The lifetime corresponding to the sum of all the spectra is given as a full horizontal line.

In order to keep acquisition time low, strong sources were used (up to 50 µCi). Such strong sources increase the background, but test measurements with weaker sources showed that this had little impact on the results. As a second measure, the channel width of the time-to-analog converter was set to 50 ps to get more counts into the individual channels, see Fig. 2a. Moreover, the detectors were placed closely together. Whenever no cryostat was used the distance could be very small (a few mm). With all these measures, lifetime values could be obtained in 0.75 min (see Fig. 5c). This corresponds to spectra with just 1/3 of the counts in Fig. 2a). However, for the sake of a lower statistical error an acquisition time of 2 min was chosen in most cases. Some of the characteristics of the three lifetime spectrometers used are summarised in Table. 1.

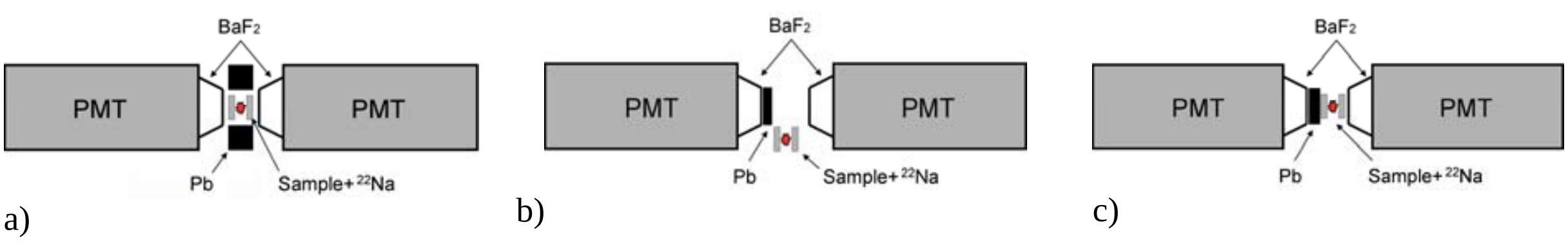

Fig. 3. Schematic diagram of 3 different detector/sample arrangements. The detector for the 511keV photon is on the right.

Table 1. Some characteristics of the PALS spectrometers used.

| | | | |
|---|---|---|---|
| location of spectrometer | CSIRO/Melbourne | Halle | Berlin |
| scintillator material | Plastic | Plastic | $BaF_2$ |
| source | $^{22}$NaCl in Ti foil (50 µCi) | $^{22}Na_2CO_3$, Kapton foil (49.5 µCi) | $^{22}Na_2CO_3$ in Kapton foil (26 µCi) |
| source contribution | 5% 0.38 ns + 0.25% 2 ns | 10.6% 0.39 ns + 0.33% 2.33 ns | 11.8% 0.387 ns + ≈0.7% 3.5 ns (Al standard)<br>[12.7% 0.394 ns + ≈0.7% 3.7 ns (Si standard) used for test purposes only] |
| detector resolution | 0.245 ns | 0.220 ns | 0.255 – 0.260 ns |
| detector/sample arrangement | coaxial, sample on axis | coaxial, sample on axis | coaxial, sample on axis, 5 mm lead shielding of 1275 keV detector |
| count rate in cps | 660 (Fig. 5a) / 1250 (Fig. 5c) | 970 (Fig. 5a,b) | 975 (Fig. 5a) |
| background / peak counts* | 0.1% / 0.15 % | 0.06% | 0.03% |

* at 15,000 peak counts corresponding to ≈150,000 total counts

For the $BaF_2$ scintillators used, backscattering of γ-photons from the high energy side is known to cause problems. Therefore, we tested 3 different detector/sample (the latter containing the source) arrangements. First, an annealed Al (5N) reference sample was placed inside the opening of a lead plate placed between two co-linear detectors, as illustrated in Fig. 3a. The 0.130 ns lifetime obtained was far below the literature value of 0.160 ns due to strong backscattering. Tuning the energy window of the stop channel to a narrow window around 511 keV led to more realistic values but very low count rates ~15 cps. Arranging detector and sample in a way suggested in the literature [7][8], see Fig. 3b, improved the situation. Here, a 5-mm thick lead plate between the two detectors and an offset of the sample position suppressed backscattering and yielded a realistic lifetime of 0.163 ns but a low count rate around 100 cps. However, the spectrum distortion caused by the summation of the 1.28 MeV γ and one of the 511 keV annihilation γ in the start channel could be avoided.

The best compromise was found by moving the sample back to the common axis, see Fig. 3c, thus increasing the count rate to 1000 cps and yielding realistic lifetimes. Other than reported in Ref. [6], consistent source corrections could be obtained from both annealed Al and Si reference samples, see Table 1.

### *2.3. Doppler broadening spectroscopy (DB)*

Doppler broadening measurements were carried out with a single HP solid-state germanium photon detector (Ortec GEM Series, model F5930P4) kept at ≈–170°C using an ORTEC X-Cooler II. The scintillator crystal was 58.6 mm in diameter, 32.5 mm long and contained a 0.7 mm inactive dead layer and a 1.27 mm Al absorbing layer. The samples used were 3 mm thick (≈1 $cm^2$ in area) to ensure that all the positrons annihilated within the specimen. A sandwich configuration of two simultaneously quenched samples was chosen. The positron source (< 0.5 $mm^3$ volume) was placed into a small pit drilled into one of the samples prior to solutionising. The sample sandwich was kept at temperatures ≤–90°C during DB measurement by placing it on a cold finger in front of the detector. The source-to-detector distance was adjusted to ≈5 cm, so that the count rate was ≈1000 cps. DB spectra were stored once per minute in individual data files.

Data analysis was carried out using the program MSpec (version 2.092) developed by M. Haaks. The individual spectra were corrected for peak shifts (using the 511 keV peak as a reference), then added up, after which a high momentum analysis (HMA) was carried out involving subtraction of various background components [9]. HMA requires an empirical choice of two energy windows, within which the background corrections are carried out. This selection was done intuitively by minimising the asymmetry of the annihilation peak.

### *2.4. Temperature programs*

Solutionising for 30 min at 535°C (or 540°C) and quenching into ice water was the first step in all the experiments. Drying and sandwiching two samples and a positron source took about 90–160 s, during which the temperature equilibrated. The samples were then either cooled to –50°C for static measurements (a) or were immediately characterised dynamically (b). Artificial ageing steps without (c) or with (d) natural pre-ageing (before sample assembly) were carried out to study the interactions between the various ageing stages.

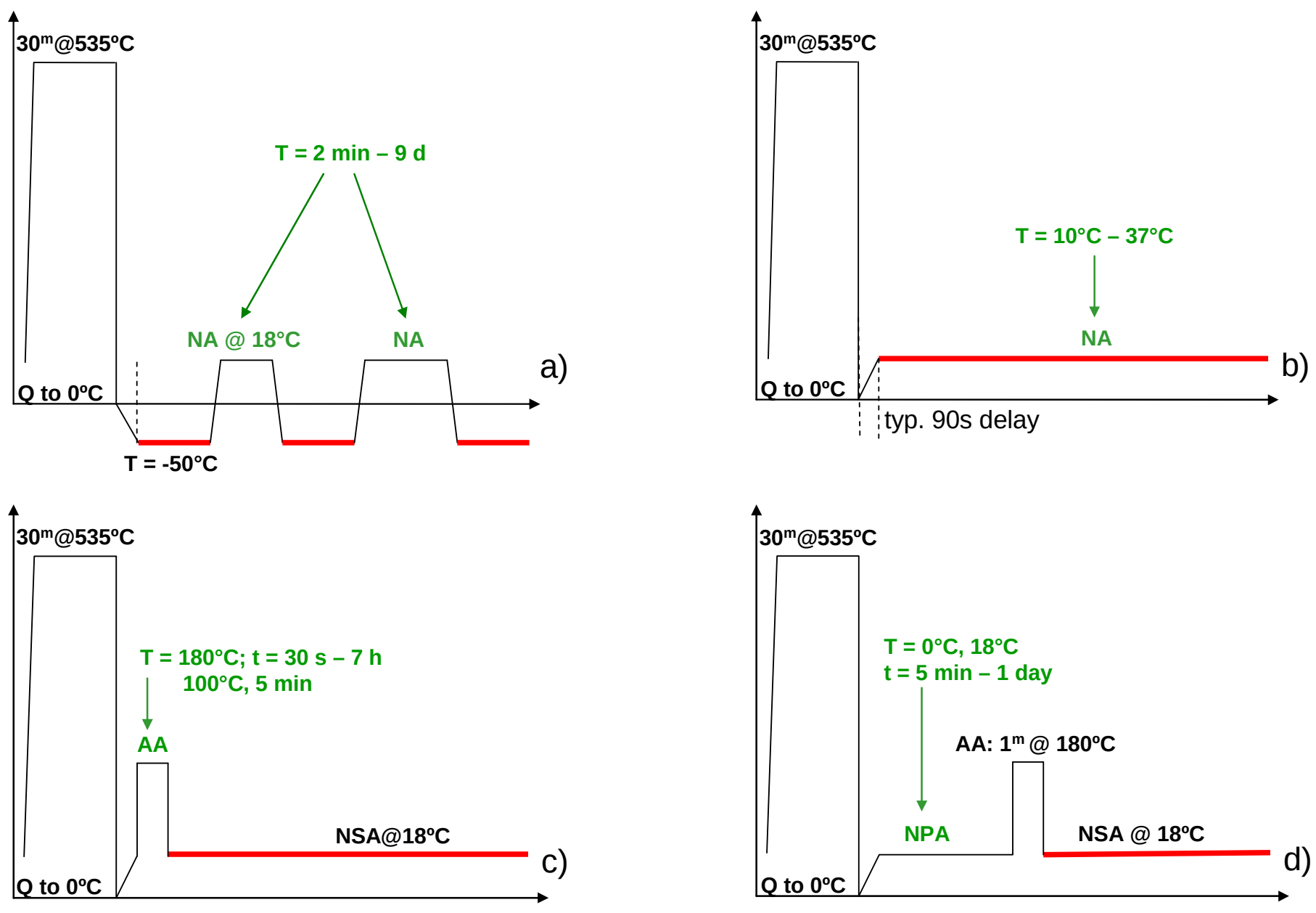

Fig. 4. Experiment types employed in the present study. Red: continuous PALS or DB measurement, Green: parameters varied. a) Static PAS measurements at low temperature where microstructure is stable. Ageing treatments are carried out between individual measurements. b) Dynamic isothermal PAS measurements after quenching, c) same as b), but with additional artificial ageing step after which natural (secondary) ageing (NSA) is carried out, d) same as c) but with additional natural pre-ageing (NPA) step.

## 3. Results

### *3.1. Isothermal PALS experiments according to Fig. 4b*

Immediately after quenching the alloys, the average positron lifetime started to decrease from an initial value of up to 0.250 ns down to typically 0.215 ns. The decrease takes between 40 and 160 min, depending on both alloy composition and temperature. From the temperature dependence of the decrease an activation energy of 85 kJ/mol has been derived [6]. The lifetime re-increases and runs through a maximum, after which it finally decreases asymptotically. Fig. 5a, shows this typical course for alloy 4-4 measured with all three spectrometers. The CSIRO measurements exhibit an offset to lower values, most likely because some annihilation in the Ti foil around the $^{22}$NaCl source is contained in the signal. The measurement in Halle might have suffered from a slight increase in laboratory temperature during the measurement since the positions of the minimum/maximum deviate. Fig. 5b demonstrates that it is acceptable to re-bin data of the decay spectra – i.e. to increase the spacing of points in Fig. 2a that equals the channel width $t_{TAC}$ of the time-to-amplitude converter – in order to reduce the acquisition time of each spectrum as the only difference found is a small offset. Fig. 5c demonstrates that in alloys higher in solute the initial lifetime values are lower and that there is a stage of constant lifetime for the first ≈10 min. Moreover, various averaging procedures are compared. Fig. 5d demonstrates that various levels of impurities in alloys similar to 4-10 have a slight influence on the kinetics of NA, but this influence is less pronounced than that known from artificial ageing.

### *3.2. PALS experiment with prior artificial ageing (AA) according to Fig. 4c*

If directly after quenching an alloy 6-8 is exposed to an ageing treatment at 180°C, after which the lifetime is continuously measured, we observe lifetimes that are much lower than without such treatment (0.204–0.210 vs. 0.235 ns). The measured low values are given in Fig. 6a. There is a clear dependence on AA time with a minimum in the range of an hour. This behaviour has been reported previously in a similar case [11]. During PALS measurement at ‘room temperature’, the lifetime increases for $t_{AA} \leq 5$ min, i.e. natural secondary ageing (NSA) has a notable influence. For 1 min of AA this will be discussed in Sec. 3.3. The range of lifetime increase during NSA is indicated by arrows in Fig. 6a.

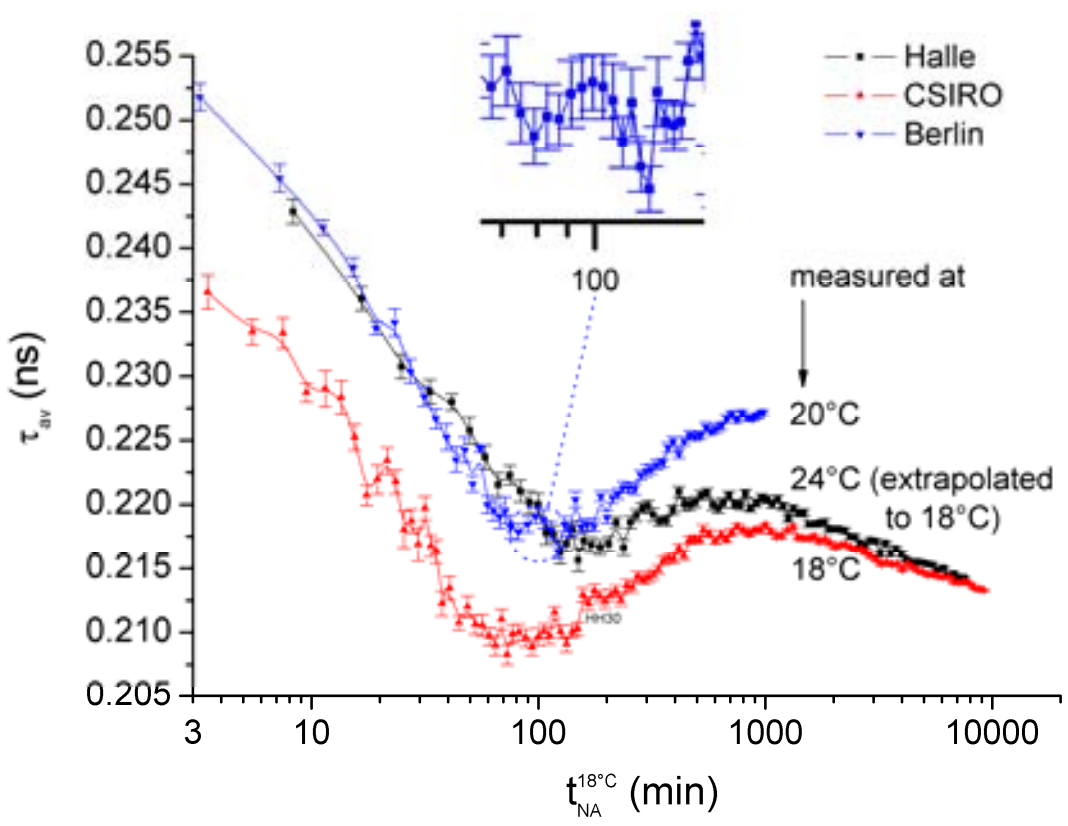

a)

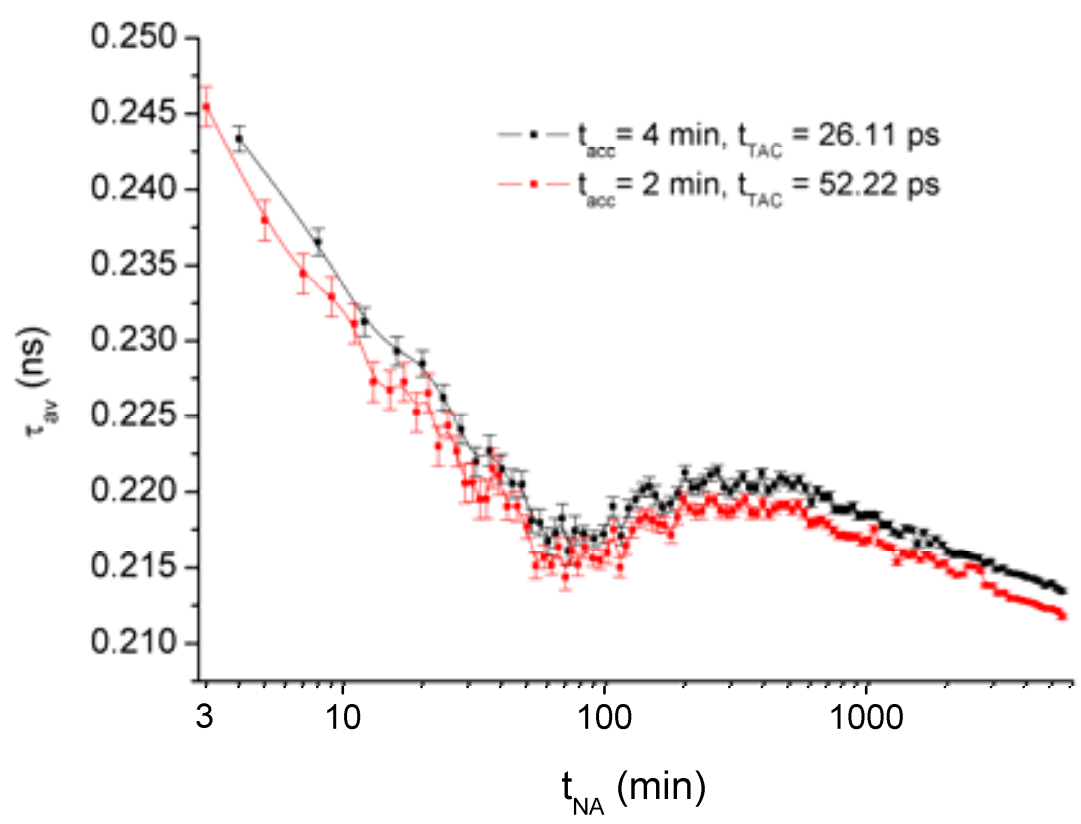

b)

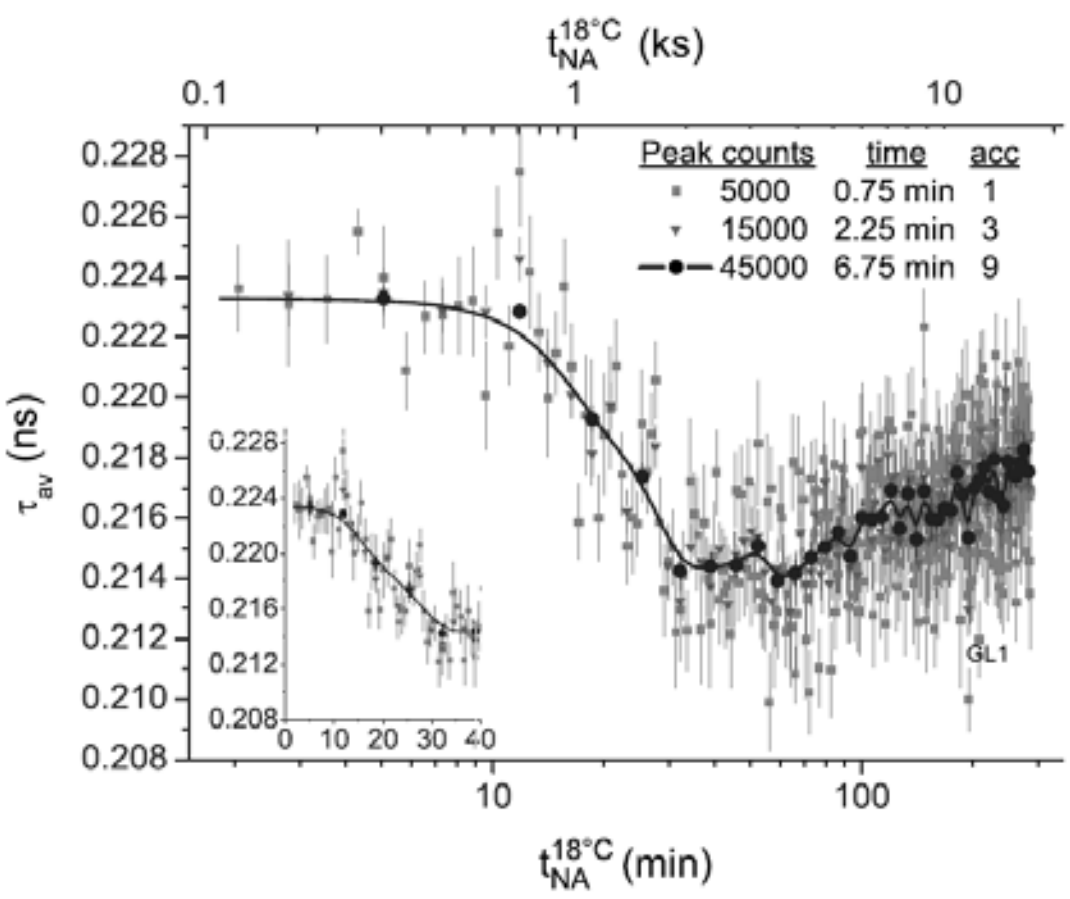

c)

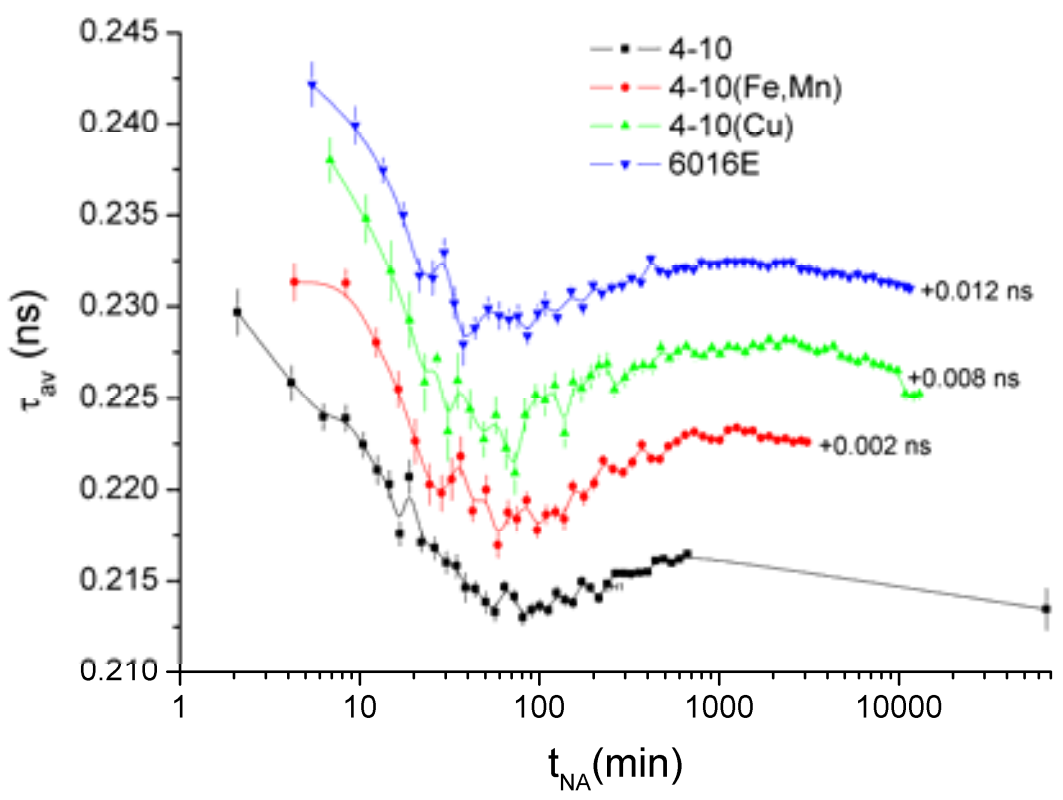

d)

Fig. 5. a) Development of average positron lifetime in alloy 4-4 at 18°C, directly after solutionising and quenching, see Fig. 4b. Comparison of different spectrometers. Curve measured at 24°C has been adjusted to 18°C by stretching the time axis by a Boltzmann factor (2.07) based on these two temperatures and an activation energy of 87 kJ/mol. Inset shows magnification of the minimum of one of the curves. b) Comparison of lifetime analysis based on data accumulated for a given time $t_{acc}$ and a bin width $t_{TAC}$ of the time-to-amplitude converter (T=24°C). c) Average positron lifetime in alloy 8-6. Three channel widths $t_{acc}$ are compared: 45 s, 2.25 min and 6.75 min, while $t_{TAC}$=50 ps [6]. d) Average positron lifetime in four alloys including Si-rich alloy 4-10 and two variants containing impurities of (Fe,Mn) or Cu and a related industrial alloy 6016 (note that an offset has been introduced to separate the individual curves).

### *3.3. PALS experiment with combined natural pre-ageing (NPA) and artificial ageing (AA) according to Fig. 4d*

As mentioned in Sec. 3.2, AA after quenching leads to a rapid reduction of the average positron lifetime, see Fig. 6a, the final value depending on AA time. Natural secondary ageing (NSA) after this leads to a re-increase as indicated by the arrows in Fig. 6a. The extent of this lifetime increase during NSA depends on the alloy and AA time, but also on a natural pre-ageing (NPA) step before AA, see Fig. 6b and 6c. The longer pre-ageing is, the smaller the lifetime change. 10 min at 18°C are sufficient to eliminate the effect, while it is still present after 40 min at 0°C. These 10 min at 18°C roughly match the period in which the negative effect of NA on age hardening is established.

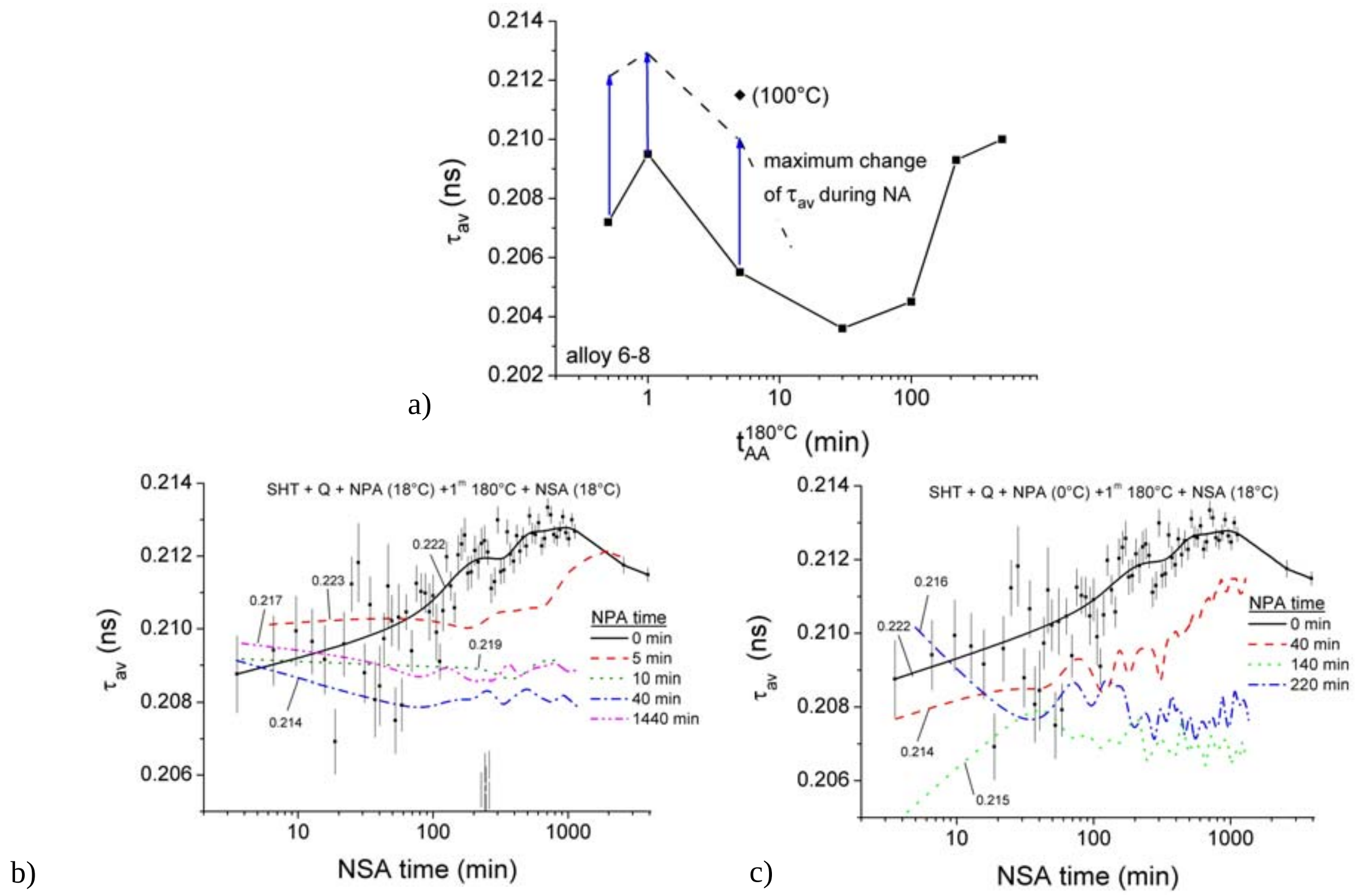

Fig. 6. a) Positron lifetime in alloy 6-8 immediately after artificial ageing at 180°C for different times (one value for 100°C given). The arrows indicate how much this lifetime changes during ensuing NSA at 18°C after AA (value for $t_{AA}$=1 min can be read from (b)). b) Positron lifetimes during natural secondary ageing (NSA) at 18°C after NPA at 18°C for different times and AA for 1 min at 180°C. Individual data points are shown for one of the curves only. c) same as b) but NPA at 0°C. The lifetimes immediately before AA are given for each curve (taken from [10]). Note that curves corresponding to 0 min of NSA are the same in b) and c).

### *3.4. Doppler broadening (DB) measurements on naturally aged alloys according to Fig. 4a*

Six DB profiles of alloy 6-8 after quenching and NA for different times are shown in Fig. 7a as ratios to pure Al. “0” min means that directly after ice water quenching and drying the sample was quenched into liquid nitrogen, so that the delay at ‘room temperature’ was minimal. The Doppler profiles evolve with NA time in the following way (see arrows): 1) The value at p=0 (p is electron momentum) increases (and with this the S parameter). 2) The hump around p=8 disappears and develops into a shoulder. 3) In the high-momentum region there is a trend of decreasing ratio, although some lines cross in the region of lower count rate. Ratio data is also shown for annealed Mg (3N) and Si (5N) and for quenched Al (5N).

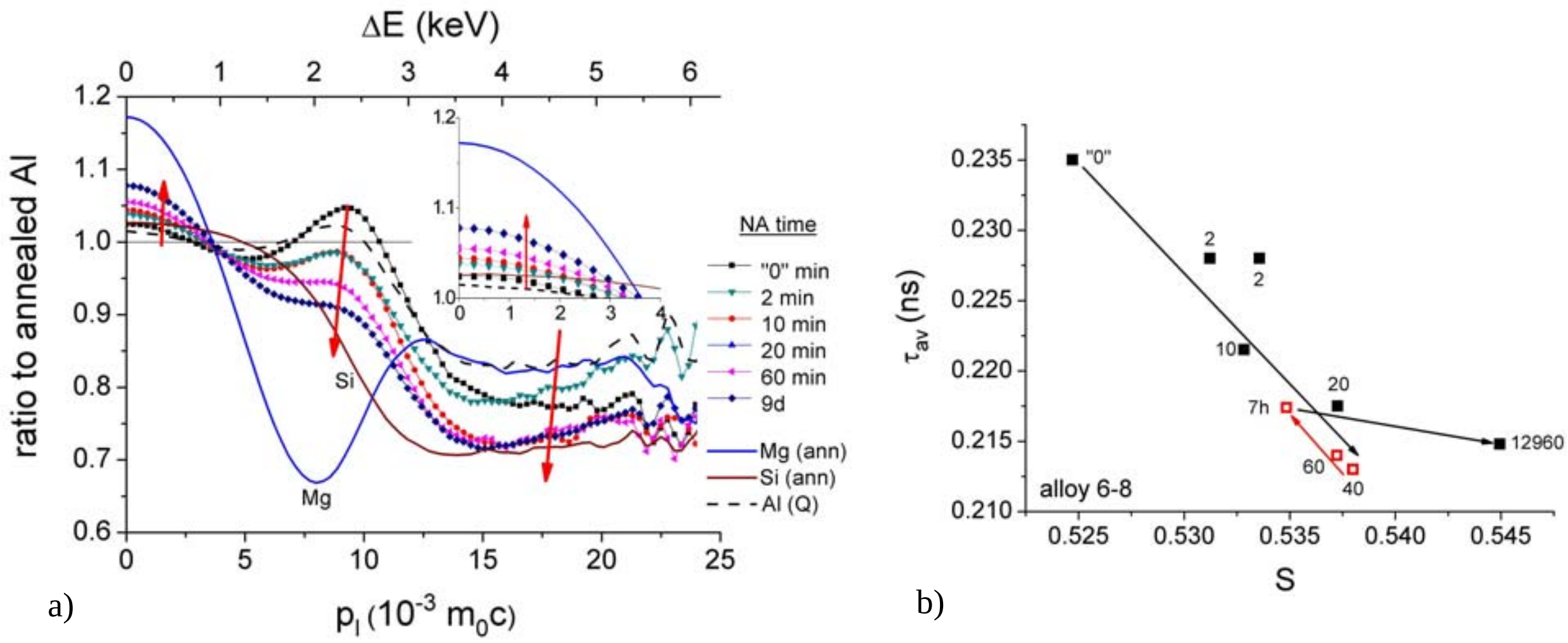

Fig. 7. a) Doppler broadening ratio plots of alloy 6-8 in various states of NA with respect to annealed pure Al (5N). The lower axis shows longitudinal electron momentum, the upper the difference to the centre of the 511 keV annihilation line. Curves for quenched ('Q') Al and annealed ('ann') Mg and Si are also given. b) Positron lifetime given as a function of S parameter. The numbers denote NA time in min and open symbols refer to the stage in which the lifetime is increasing.

## 4. Discussion

During solutionising ('*stage S*') a high density of free vacancies (≈$10^{-4}$) would give rise to saturated positron trapping and to a single positron lifetime of ≈0.250 ns. This has been measured directly for pure Al [12].

2 min after quenching, initial lifetimes in alloys low in solute (<0.75%) range between 0.240–0.255 ns and indicate that quenching was fast enough to largely preserve this situation. According to still preliminary theoretical work [13] the formation of vacancy-solute complexes would not change the lifetime very much, so that these might have formed. This is very likely since their formation prevents the vacancies from quickly diffusing to sinks as it happens in very pure aluminium. In alloys with higher solute contents (>1.4%) we measure lower initial lifetimes 2 min after quenching, see Fig. 5c. The same applies to low-solute alloys that were not quenched in ice water but into water at 'room temperature' or in air. This allows us to conclude that already during quenching ('*stage Q*') positron traps are formed in which positrons annihilate faster than in free vacancies. Alloys with >1.4% Mg+Si and especially those high in Mg (alloy 10-4, initial lifetime=0.214 ns [6]) are quench-sensitive enough to show this effect during ice water quenching, whereas low-solute alloys exhibit it only during slower quenching. In all those cases, a short time spent at elevated temperatures before reaching temperatures low enough to freeze these processes is sufficient to create positron traps.

Exposing quenched alloys to 180°C directly after quenching (*stage AA*) has a similar effect: the lifetime decreases by more than 0.015 ns during holding for just 30 s, see Fig. 6a. It is very likely that in both cases (during quenching, during holding at 180°C) similar processes take places. Our analysis of spectra after ageing at 180°C show that a possible reduced bulk component contributes less than <10% to total intensity (>90% due to traps with 0.200-0.210 ns). Staab et al. find no bulk component in similar alloys 6013 [11]. A recent study at the Universities of Rostock and Halle of alloy 6005 quenched at different rates revealed that above 10 K/s annihilation in vacancy-related defects (0.225 ns) contributes ≈80% to the decay spectrum [14]. Slower quenching leads to increasing bulk contributions. We found that alloy 4-4 after 2 h of annealing at 400°C exhibits pure bulk annihilation (0.160 ns). From this evidence one has to conclude that the new positron traps formed during quenching and short annealing at 180°C are predominantly coherent clusters or precipitates of solute atoms, both vacancy free. However, a more thorough analysis of the lifetime spectra during such annealing seems necessary to quantify concurrent annihilation in vacancies and the bulk.

During NA, there is a period of almost constant or even slightly increasing lifetime for some of the alloys, called *stage I*, which can currently no be explained, see Fig. 5a and [6]. It is known that the negative effect of NA on AA develops within this period which might be more than a coincidence [3]. The pronounced lifetime decrease by up to 0.030 ns within the first hour (*stage II*) must be caused by trapping of positrons into newly generated structures in which they annihilate faster than in vacancies or vacancy-solute complexes. A gradual increase of bulk annihilation (≤0.160 ns) in this stage caused by the loss of vacancies, which would explain the observed decrease, could not be verified by measurements. As with the lifetime decrease in stages Q and AA, the formation or further growth of vacancy-free positron traps that compete with annihilation in vacancy-related sites must be held responsible. According to this scenario, trapping would be saturated but a shift from vacancy to cluster-related trapping would cause the decrease at all times. Evidence for this is provided by recent work by Klobes, who carried out low-temperature PALS measurements on the same alloy and found that 30 min after quenching the positron lifetime at 50 K is up to 0.016 ns lower than at 230 K, indicating that shallow traps exist [15]. Such traps do not contain vacancies but are agglomerates of Si or Mg (or both) which are coherently embedded in the Al matrix. In such objects positrons can have lifetimes lower than in vacancies but higher than in bulk Al and therefore do not give rise to a second observable component in our lifetime spectra due to the limited resolution of the spectrometers used. Precision measurements with high-resolution spectrometers would be helpful to verify this point.

The re-increase of lifetime after a minimum (*stage III*) occurs only in alloys containing Mg. Moreover, Mg accelerates this increase. It is not known whether the lifetime increase is caused by Mg atoms via their interactions with positrons or whether Mg influences the thermodynamics of cluster growth and influences positrons only indirectly. The final stage is characterised by a small decrease (*stage IV*), perhaps due to cluster coarsening or ordering.

The DB data shown in Fig. 7a suggest that immediately after quenching the signal is dominated by the hump around p=8 which is typical for vacancies in pure aluminium. The occurrence of this hump is confirmed by Ref. [16], whereas others merely find a shoulder here [15]. The feature is actually more pronounced in the alloy than in the quenched pure Al specimen, probably because the vacancies migrate faster to sinks in pure Al than in the alloy. NA causes the DB curves to move down towards the values of both pure Si and Mg in this region. The high momentum region (p>15) indicates a dominant influence of Si as the curves come much closer to Si than to Mg here. (Note: The ratio curve for Mg quenched from 540°C (not shown) is even higher than that for annealed Mg in the high-momentum region. Si should not contain vacancies in neither case, annealed or quenched). Fig. 7b relates the development of the S parameter and positron lifetime. Clearly both are correlated which is due to the fact that both are related to defect density.

The small hump that can be observed in the lifetime spectra near the minimum of NA curves, see inset of Fig. 5a, was taken as noise in previous work [6] but actually occurs in many of the measurements. This is an example not only for the complexity of the NA process in which various processes compete with each other but also for the necessity for further precise measurements.

The situation is more complicated when multistage ageing procedures are applied. One minute of annealing at 180°C directly after quenching brings down the average positron lifetime from about 0.222 ns to 0.209 ns, see Fig. 6a,b. The trend is the same as for NA, just that the lifetime decrease during AA is much faster. The positron lifetime spectra now most likely contain a component related to a coherent cluster of solute atoms, perhaps with a small bulk component that, however, cannot be resolved. NA after 1 min of AA leads to a lifetime increase by 0.004 ns within ≈500 min. Such a lifetime increase due to secondary ageing has also been observed for Al-Cu-Mg alloys [17,18]. The clusters formed at 180°C will remain stable at 'room temperature' so that the observed increase of positron lifetime is either due to a continued growth and modification of these or the formation of additional trapping sites in the matrix around these clusters. In Ref. 17, the latter viewpoint is adopted and the formation of new traps in addition to the ones formed during AA ascribed to the action of Mg in a first and to Cu in a second stage.

A natural pre-ageing (NPA) step before AA as short as 10 min at 18°C or in the order of 60 – 140 min at 0°C eliminates the increase during NSA. Clustering taking place during NPA modifies the precipitation kinetics during subsequent AA. For long AA times this is known: fewer and coarser precipitates are formed and the hardness increase is limited ('negative effect') [3], but the structure after 1 min of AA is not known. The experiments show that in this state no more positron traps can be formed that increase the average lifetime.

## 5. Conclusions

Positron lifetime measurements can be carried out with a time resolution of ≤1 min as long as the spectra can be described by one lifetime component. This has to be checked by independent low-temperature and high-resolution measurements. Lifetime variations take place in up to 4 stages which correspond to the time evolution of electrical resistivity, hardness and heat release. Positron lifetime spectroscopy was shown to be useful due to its very high sensitivity and reproducibility, but suffers from the difficulty in explaining the observed lifetime changes in all cases. Further progress in interpretation requires input from ab-initio calculations and more precise experiments.

## Acknowledgements

We would like to thank Prof. J. Hirsch (Hydro Aluminium) for providing samples, Prof. R. Krause-Rehberg (University of Halle) for his help and advise with PAS, Prof. K. Maier and Dr. B. Klobes (University Bonn) for borrowing and explaining us their spectrometer and Dr. M. Haaks for providing a $^{68}$Ge source and teaching us high-momentum analysis.

## References

[1] M.A. van Huis, J.H. Chen, M.H.F. Sluiter, H.W. Zandbergen, Acta Mater. 55, 2183 (2007).
[2] C.S.T. Chang, J. Banhart, Met. Mater. Trans. A 42(7), 1960 (2011).
[3] J. Banhart, M.D.H Lay, C.S.T. Chang, A.J. Hill, Adv. Eng. Mater. 12, 559 (2010).
[4] M. Torsæter, Thesis, NTNU Trondheim (2011).
[5] A. Dupasquier, G. Kögel, A. Somoza, Acta.Mater. 52, 4707 (2004).
[6] J. Banhart, M.D.H Lay, C.S.T. Chang, A.J. Hill, Phys.Rev. B 83, 014101 (2011).
[7] J. De Vries and F.E.T. Kelling. Nucl. Instr. Meth. Phys. Res. A 262, 385 (1987).
[8] Li Hui, Shao Yundong, Zhou Kai, Pang Jingbiao and Wang Zhu. Nucl. Instr. Meth. Phys. Res. A 625, 29 (2011).
[9] M. Haaks, T.E.M. Staab, K. Maier, Nucl. Instr. Meth. Phys. Res. A 569, 829 (2006).
[10] J. Banhart, C.S.T. Chang, Z. Liang, N. Wanderka, M.D.H. Lay, A.J. Hill, The kinetics of NA in 6000 alloy - a multi-method approach, 12th International Conference on Aluminium Alloys (ICAA12), 05-09 September 2010, Yokohama, Japan, Proceedings, Editors: S. Kumai, O. Umezawa, Y. Takayama, T. Tsuchida, T. Sato, The Japan Institute of Light Metals, Tokyo, pp. 381 (2010).
[11] T.E.M. Staab, R. Krause-Rehberg, U. Hornauer, E. Zschech, J. Mater. Sci. 41, 1059 (2006).
[12] H.-E. Schäfer, phys. stat. sol. (a) 102, 47 (1987).
[13] B. Korff and M. Offenberger, private communication.
[14] M. Wichmann, Leerstellenkonzentration in EN AW-6005A mit Positronenannihilation, Report, University of Rostock (2011).
[15] B. Klobes, Thesis, University of Bonn (2010).
[16] A. Dupasquier, R. Ferragut, M.M. Iglesias, M. Massazza, G. Riontino, P. Mengucci, G. Barucca, C.E. Macchi, A. Somoza, Phil. Mag. 87, 3297 (2007) .
[17] A. Somoza, A. Dupasquier, I.J. Polmear, P. Folegati, R. Ferragut, Phys. Rev. B61, 14454 (2000).
[18] M. Massazza, G. Riontino, A. Dupasquier, R. Ferragut, A. Somoza, P. Folegati, Phil. Mag. 82, 495 (2002).